\documentclass[12pt]{article}
\usepackage{cite}
\usepackage{amsmath, amsthm, amssymb,slashed,url}
\usepackage{ifpdf}
\ifpdf
  \usepackage[pdftex]{graphicx}
  \usepackage{epstopdf}
\else 
  \usepackage[dvips]{graphicx}
\fi
\parskip=\baselineskip
\def\Bbb{\mathbb}
\def\Tr{{\rm Tr}}
\def\16{{\bf 16}}
\def\1{{\bf 1}}
\def\2{{\bf 2}}
\def\3{{\bf 3}}
\def\4{{\bf 4}}
\def\bar{\overline}
\def\ovg{\overline g}

\def\R{{\Bbb{R}}}

\def\h{\widehat}
\font\teneurm=eurm10 \font\seveneurm=eurm7 \font\fiveeurm=eurm5
\newfam\eurmfam
\textfont\eurmfam=\teneurm \scriptfont\eurmfam=\seveneurm
\scriptscriptfont\eurmfam=\fiveeurm

\font\teneusm=eusm10 \font\seveneusm=eusm7 \font\fiveeusm=eusm5
\newfam\eusmfam
\textfont\eusmfam=\teneusm \scriptfont\eusmfam=\seveneusm
\scriptscriptfont\eusmfam=\fiveeusm

\font\tencmmib=cmmib10 \skewchar\tencmmib='177
\font\sevencmmib=cmmib7 \skewchar\sevencmmib='177
\font\fivecmmib=cmmib5 \skewchar\fivecmmib='177
\newfam\cmmibfam
\textfont\cmmibfam=\tencmmib \scriptfont\cmmibfam=\sevencmmib
\scriptscriptfont\cmmibfam=\fivecmmib

\numberwithin{equation}{section}

\def\d{\mathrm d}

\def\D{{L}}
\def\i{{\mathrm i}}
\def\A{{A_0}}
\def\a{{\sf a}}
\def\K{{\sf K}}
\def\h{\widehat}
\def\bar{\overline}
\def\be{\begin{equation}}
\def\ee{\end{equation}}
\begin{document}
\begin{titlepage}
\begin{flushright}

\end{flushright}
\vskip 1.5in
\begin{center}
{\bf\Large{A Note On Boundary Conditions \\ \vskip.5cm In Euclidean Gravity}}
\vskip
0.5cm {Edward Witten} \vskip 0.05in {\small{ \textit{School of
Natural Sciences, Institute for Advanced Study}\vskip -.4cm
{\textit{Einstein Drive, Princeton, NJ 08540 USA}}}
}
\end{center}
\vskip 0.5in
\baselineskip 16pt
\abstract{We review what is known about boundary conditions in General Relativity on a spacetime of Euclidean signature.
The obvious Dirichlet boundary condition, in which one
specifies the boundary geometry, is actually not elliptic and in general does not lead to a well-defined perturbation theory.  
It is better-behaved if the extrinsic curvature of the boundary is suitably constrained, for instance if it is positive- or negative-definite.
A different boundary condition, in which one specifies the conformal geometry of the boundary and the trace of the extrinsic curvature, 
is elliptic and always leads formally   to a satisfactory perturbation theory.  
These facts might have interesting implications for semiclassical approaches to quantum gravity.}
\date{April, 2018}
\end{titlepage}
\def\Hom{\mathrm{Hom}}

\tableofcontents
\section{Introduction}

The goal of this note is to make accessible some  basic properties of boundary conditions in Euclidean gravity \cite{Anderson,Anderson2,AE,Hamilton}.   
The facts described here are not new.   The motivation for presenting this material is that it may have applications to semiclassical quantization of gravity.  
There is an extensive
literature on this subject, a small selection being \cite{GH,GP,CD,GHP,HH,Polch,AEK,BB}.    

In Euclidean signature, let $X$ be a $D=d+1$-dimensional Riemannian manifold with boundary a $d$-manifold
$M$.   The most obvious boundary condition in Euclidean quantum gravity is to specify the geometry of $M$ (that is, its Riemannian metric) and
integrate over all metrics on $X$ that are consistent with this boundary geometry.   Another and equally simple boundary
condition is to specify the conformal geometry of $M$ and the trace $\K$ of its extrinsic curvature or second fundamental form (which is roughly the normal derivative
of the metric along $M$).     We will call these the Dirichlet and conformal boundary conditions, respectively.
The main point that we aim to explain is that in general the Dirichlet boundary condition does not lead to a sensible perturbation expansion, while
the conformal boundary condition always does lead formally (that is, modulo the usual ultraviolet divergences) to a sensible perturbation expansion about
any given classical solution.  Technically, the conformal boundary condition is elliptic, ensuring its good behavior, and the Dirichlet boundary condition is not.

The conformal boundary condition is natural in the context of using the conformal structure of an initial value surface (rather than the full Riemannian
geometry of that surface) and the trace of the extrinsic curvature as a maximal set of commuting variables.  This idea has a long history \cite{York}.

Although the Dirichlet boundary condition is ill-behaved in general, one can show,  by using the ellipticity of a certain alternate boundary condition that is similar
to the conformal one,
that the Dirichlet boundary condition is well-behaved at least in some respects 
if the extrinsic curvature of the boundary satisfies a certain condition, for instance if it is
 positive- or negative definite.    (The precise statement here is a little subtle, as we discuss momentarily.)

In section \ref{two}, we review the concept of an elliptic boundary condition, and then in section \ref{three}, 
we explore this concept in the context of 
gravity.  In gauge theory, there is a very natural elliptic boundary condition in which the boundary value of the gauge
field is specified.  Gravity is different, basically because of the second order nature of the Hamiltonian constraint equation.  As 
already remarked, one can get an elliptic boundary condition by specifying the conformal class of the boundary metric and the trace of the
second fundamental form, but not by specifying the boundary metric.

To avoid confusion, we should stress that ellipticity of a boundary condition in General Relativity
does not guarantee either existence or uniqueness of a solution of the Einstein
equations with specified boundary data. Nor does it have anything to do with positivity of the operator that governs gravitational perturbations.
 Ellipticity does guarantee  the properties that are needed to construct perturbation theory.   It ensures that the gauge-fixed Einstein equations (on a compact
 manifold) have only finitely many zero-modes, and that modulo these zero-modes, the linearized Einstein equations have a propagator with the usual
 properties.\footnote{\label{zolgo} Technically, ellipticity is the
right condition on a boundary condition to guarantee these properties if gauge-fixing is carried out in the standard way,
so that  the gauge-fixed action for metric fluctuations
has a nondegenerate second order kinetic energy.  In some other approaches to gauge-fixing, one would 
encounter a slight generalization of the notion of ellipticity.   See the remark at the end
of section \ref{ym}.}   
    Mathematically, a differential operator $L$ with a finite-dimensional kernel and cokernel\footnote{The kernel of an operator $L$ is the space of solutions of $Lu=0$.  To define the cokernel,
 consider the more general equation $Lu=f$ with a source $f$.  The cokernel is the space of all $f$'s modulo those for which the equation can be solved.
 If $L$ is self-adjoint -- as are the linearized Einstein equations with some boundary conditions -- the condition of finite-dimensionality of the cokernel
 is redundant as 
  there is a natural isomorphism between the kernel and the cokernel.} is said to be Fredholm; and a propagator defined
 after removing a finite-dimensional space of zero-modes is called a parametrix.   
 
Quantum perturbation theory is usually constructed by expanding around a classical solution. 
Generically ellipticity is needed to ensure the properties that make perturbation theory possible.\footnote{Perturbation
theory also requires the absence of certain one-loop anomalies, which involve topological considerations.  
 Further issues arise if a theory has gauge symmetries that do not allow any useful regularization.
This is relatively uncommon.}     However, for the specific
case of General Relativity, one can show that if the extrinsic curvature of the boundary
 is positive- or negative-definite (and under somewhat more general conditions), the linearized Einstein equations with Dirichlet boundary conditions are Fredholm
even though not elliptic.  It seems plausible that they admit a parametrix, though this does not seem to be rigorously known. So this may be a case in which perturbation
theory is possible with a boundary condition that is not elliptic.

A lecture by the author on issues possibly related to what is
described here can be found in \cite{Talk}.

It is a pleasure to dedicate this article to Roman Jackiw on the occasion of his birthday.
Roman has introduced many important ideas in physics.   The triangle anomaly was a milestone
and a turning point in the understanding of the strong interactions.   His work on zero-modes
of fermions in the field of a soliton or instanton has been very important in both particle physics
and condensed matter physics.   Also important for both relativistic physics and condensed matter
physics has been Roman's work on topologically massive gauge theories in three spacetime dimensions.
All these things have in particular been important for my own work.
Finally, since this article is largely concerned with gravity,
I think I should mention Roman's work with Curt Callan and Sidney Coleman \cite{CCJ}   in which they constructed
a new ``improved'' energy-momentum tensor with a softer trace, more natural for the coupling of a quantum
field theory to gravity

\section{Background}\label{two}

\subsection{Elliptic Boundary Conditions}\label{ellbc}

First let us recall the definition of an elliptic differential operator.  The symbol of a differential operator is defined, roughly speaking,
by replacing derivatives $-\i \partial/ \partial x^\mu$ by momenta $p_\mu$.    To be more exact, if $\D$ is a differential operator of order
$n$ on a manifold $X$, then its ``leading symbol''  is defined by making the substitution $-\i \partial/\partial x^\mu\to p_\mu$ in the terms of order $n$
and dropping terms of order less than $n$.   Thus the leading symbol of $\D$ is, for each point $x\in X$, a polynomial $\sigma_x(p)$ in 
$p$ that
is homogeneous of degree $n$.   For example, the leading symbol of the Laplacian
\be\label{delfo}\Delta=-g^{\mu\nu}D_\mu D_\nu \ee
is $\sigma_x(p)=g^{\mu\nu}(x) p_\mu p_\nu$.   Likewise, the leading symbol of the Dirac operator
\be\label{welfo}\i\slashed{D}=\i \Gamma^\mu D_\mu \ee
(where $\Gamma^\mu$ are gamma matrices) is $\sigma_x(p)=\slashed{p}=\Gamma\cdot p$.

If $\D$ has order $n$, then the leading symbol $\sigma_x(p)$ 
is naturally understood as a function -- or more generally, as in the Dirac case, a matrix-valued function -- 
on the cotangent bundle $T^*X$, homogeneous
of degree $n$ on the fibers.    $\D$ is called ``elliptic'' if for all $x\in X$ and all real nonzero $p$, $\sigma_x(p)$ is invertible.
For example, the leading symbols $\sigma_x(p)$ for the Laplacian and the Dirac operator have this property so these are elliptic operators.

An elliptic operator $\D$ on a compact manifold is Fredholm.   If $\D$ maps the space of sections of some vector bundle $E$ to itself (so that the eigenvalue
problem $\D\psi=\lambda\psi$ makes sense and one can define the spectrum of $\D$), then $\D$ has a discrete spectrum.  In particular, this is so if $\D$ is
self-adjoint, a common case.  
The  eigenvalues of an elliptic operator tend to infinity\footnote{Here ``infinity'' means
$+\infty$ if $\sigma_x(p)$ is hermitian and positive-definite -- as for the Laplacian -- or $\pm \infty$ if $\sigma_x(p)$ is hermitian but
not positive-definite, as for the Dirac operator.  If $\sigma_x(p)$ is invertible but not hermitian,  then   $\D$ still has a discrete spectrum but its
eigenvalues are not necessarily real, and can tend to infinity in any direction in the complex plane in which an eigenvalue of
$\sigma_x(p)$ can tend to infinity.}  in a simple way that can be described semiclassically in terms of 
$\sigma_x(p)$.  In particular, the space of zero-modes is at most finite-dimensional.  $\D$ can be inverted, on a subspace transverse
to the zero-modes, by a Green's function $G(x,x')$ that is regular for $x\not=x'$, and whose singularities for $x\to x'$ are controlled
in the standard way by an operator product expansion.  (In particular, the leading singularity for $x\to x'$ depends only on $\sigma_x(p)$.)  Such a Green's function defined
after removing zero-modes is 
called a parametrix; physically, it is used as a propagator in constructing perturbation theory.  Ellipticity, along with some topological considerations involving anomalies, also makes it possible to define a ``determinant'' of $\D$
(or in appropriate circumstances, a Pfaffian) with standard properties.  
In short, ellipticity guarantees the properties that are needed in constructing perturbation theory.\footnote{\label{delfic} The definitions just given are adequate for most
purposes, but in general  one has to
take into account different scaling weights of different fields in defining what one means by the leading symbol $\sigma_x(p)$.   The definition 
of $\sigma_x(p)$ is modified, but a satisfactory theory still requires that $\sigma_x(p) $ should be invertible for nonzero real $p$.  This situation can arise
in gauge-fixing of Yang-Mills theory or gravity if one does not integrate out the auxiliary field.  See the concluding remarks of section \ref{ym}.}

Ellipticity is an ``open'' condition on differential operators, in the sense that if $\D$ is an elliptic of order $n$, then any small perturbation of $\D$ (by terms of
order $n$ or less) does not affect ellipticity.    This is true because invertibility of $\sigma_x(p)$ is similarly an open condition, invariant under small perturbations.

Now suppose that $X$ has a boundary $M$ of dimension $d=D-1$, 
and that we are given some boundary condition for $\D$ along $M$ that reflects a boundary
condition on underlying quantum fields.  To be able to do perturbation theory in this situation, the boundary condition must
satisfy a condition that ensures that $\D$ will still have a propagator and determinant with appropriate properties.
This will always be true if the boundary condition is  ``elliptic.'' Otherwise it is typically not true.

Ellipticity of a boundary condition involves a condition that must be checked at each boundary point.
   In checking ellipticity  at a given
boundary point $x\in M$, we only care about short distance behavior, so we can approximate $X$ by a half-space
 $\R^D_+\subset \R^D$, with boundary $M\cong \R^{D-1}$, 
 and we can drop from $\D$ all terms of order less than $n$.   Moreover, ellipticity at a given boundary point $x$ only depends on $\sigma_x(p)$ for that value of $x$, so we can treat $\D$ as an operator with constant coefficients.   In other words,
if $\sigma_x(p)=\sum \sigma^{\mu_1\mu_2\dots \mu_n}p_{\mu_1}p_{\mu_2}\cdots p_{\mu_n}$, then in testing ellipticity, we can replace $\D$ with
\be\label{fluff} (-\i)^n\sum  \sigma^{\mu_1\mu_2\dots \mu_n}\frac{\partial}{\partial x^{\mu_1} } \cdots \frac{\partial}{\partial x^{\mu_n}}.\ee
For example, if $\D$ is the Laplacian of eqn. (\ref{delfo}), we can approximate it by the flat space Laplacian $\Delta_0=-\sum_{\mu=1}^D
\frac{\partial^2}{\partial x_\mu^2}$, on a half-space, say the half-space $x^D\geq 0$.  

\def\pp{p_\perp}
Let us write $\vec x =(x^1,x^2,\cdots, x^{D-1})$ for boundary coordinates and $\vec p=(p_1,p_2,\dots,p_{D-1})$ for the momentum along the boundary.  Also, let us write  $x_\perp$ for the  coordinate $x_D$ in the normal direction to the boundary, and $p_\perp$ for the momentum
in the $x_\perp$ direction.  We work on the half-space $x_\perp\geq 0$.
In the approximation of treating $\D$ as an operator with constant coefficients, the equation $\D\Phi=0$ has plane-wave
solutions $\exp(\i\vec p\cdot \vec x +\i x_\perp \pp(\vec p))$.  Here $\pp(\vec p)$ is found by solving the equation $\sigma_x(\vec p,\pp)=0$.
Let us restrict to the case that $\vec p$ is real and nonzero, so that the solution behaves as an oscillatory plane wave along the boundary.
  In this case, ellipticity of $\D$ away from the boundary means that the equation
$\sigma_x(\vec p,\pp)=0$ has no solutions for real $\pp$.    

Let us assume for the moment that  $\sigma_x(p)$ is hermitian for real $p$ (as in Yang-Mills theory and gravity with the usual gauge-fixing).  
Then, for given real nonzero $\vec p$,
the solutions of $\sigma_x(\vec p,\pp)=0$ occur in complex conjugate pairs, half with positive imaginary part of $\pp$ and half with negative
imaginary part.  The space of solutions is thus, for some nonnegative integer $s$, a $2s$-dimensional vector space $V_{2s}(\vec p)$.    The importance
of the sign of $\mathrm{Im
}\,\pp$ is simply that a plane-wave solution $\exp(\i\vec p\cdot \vec x +\i x_\perp \pp(\vec p))$ is exponentially decaying or exponentially
growing as $x_\perp$ increases, depending on the sign of $\mathrm{Im}\, \pp$.

An elliptic boundary condition is one that selects, for every real nonzero $\vec p$, a middle-dimensional subspace 
$W_s(\vec p)\subset V_{2s}(\vec p)$ of allowed solutions, with the property that 
for sufficiently large $|\vec p|$, none of the solutions in 
$W_s(\vec p)$ is exponentially
decaying with increasing $x_\perp$.  The intuitive idea is that if a 
boundary condition allows
 solutions with arbitrarily large $|\vec p|$ that are exponentially
decaying as $x_\perp$ increases, then the operator $\D$ with this boundary 
condition has too many near zero-modes that are localized
at short distances along the boundary, and cannot have a discrete spectrum (even when $X$ is compact) or a satisfactory propagator.

For a boundary condition to be local as well as elliptic means that $W_s(\vec p)$ must be defined by vanishing of an appropriate function of the fields and
their derivatives.

Let us verify that the usual Dirichlet and Neumann boundary conditions on the Laplacian are elliptic.  
The plane-wave solutions
of the Laplace equation are $\exp(\i \vec p\cdot \vec x \pm |\vec p|x_\perp)$.   Dirichlet boundary 
conditions $\phi|=0$ (where $\phi|$ denotes
the restriction of $\phi$ to $x_\perp=0$) are satisfied by the linear combination 
$\exp(\i\vec p\cdot \vec x)(e^{|\vec p|x_\perp}-e^{-|\vec p|x_\perp})$,
which is exponentially growing with $x_\perp$.   Neumann boundary conditions 
$\left.\frac{\partial \phi}{\partial x_\perp}\right|=0$ are satisfied by
the linear combination $\exp(\i\vec p\cdot \vec x)(e^{|\vec p|x_\perp}+e^{-|\vec p|x_\perp})$, which also is exponentially growing.
 So both of these boundary conditions are elliptic.
In the case of the Dirac operator, the most commonly studied boundary conditions are 
$\Gamma_D\psi|=\pm \psi$ (with some choice of the
sign).  We leave it to the reader to verify that these are elliptic boundary conditions, by showing that a solution of the Dirac equation on the half-space that
has plane wave behavior along the boundary and satisfies either of these boundary conditions is exponentially growing with increasing $x_\perp$. 

If $\D$ is self-adjoint in the absence of a boundary, then in the presence of a boundary, one frequently wants to pick a boundary
condition that is self-adjoint as well as elliptic, in other words a boundary condition that ensures that $\D$ remains self-adjoint  (as well as elliptic)
 even in the presence of a boundary.
This is a stronger condition than ellipticity alone.  For example, in the case of the Laplacian acting on a complex-valued field,  the mixed boundary condition 
$\left.\left(\frac{\partial\phi}{\partial x_\perp}-c\phi\right)\right|=0$ (which is sometimes called a Robin boundary condition)
is elliptic for any constant $c$, but it is only self-adjoint
if $c$ is real.\footnote{With Robin boundary conditions, provided $c<0$ , there are solutions
$\exp(\i \vec p\cdot \vec x -|\vec p|x_\perp)$  with $|\vec p|=-c$ that decay exponentially away from the boundary.
But for sufficiently large $|\vec p|$, there are no such solutions, so this boundary condition is elliptic.}

If $\sigma_x(p)$ is not hermitian, then solutions of $\sigma_x(\vec p,\pp)=0$ do not necessarily come in complex conjugate
pairs.  Still, an elliptic boundary condition is one that selects in the space $V(\vec p)$ of plane wave solutions with
given $\vec p$ a middle-dimensional subspace $W(\vec p)$ with the property that for sufficiently large $|\vec p|$,
no solution in $W(\vec p)$ is an exponentially decreasing function of $x_\perp$.   In general, the operator $\D$ may
not admit any local elliptic boundary condition.  The most obvious obstruction is that $V(\vec p)$ might be
odd-dimensional, and there also are further obstructions of topological nature.  In quantum field theory, the most
important example with $\sigma_x(p)$ not hermitian is the chiral Dirac operator for even $D$; it does not admit
any local elliptic boundary condition.  

We note that the condition of ellipticity -- $W(\vec p)$ does not contain any exponentially decaying solutions, for any nonzero real $\vec p$ -- has the property that if
it is true for any one boundary condition, then it is true for any sufficiently nearby boundary condition.  In this sense, ellipticity is an ``open'' condition on boundary conditions.

\subsection{Yang-Mills Theory On A Closed Manifold}\label{ym}

\def\g{{\mathfrak g}}
\def\veps{\varepsilon}
\def\Dw{\D}
\def\Dt{\D'}

Let $A$ be a Yang-Mills gauge field with gauge group $G$ and field strength $F=\d A+A\wedge A$. 
The usual action is\footnote{Here $g$ is the metric tensor of $X$.
We consider $A$ and $F$ to be real and antihermitian, so the trace on the Lie algebra is a negative-definite quadratic
form.
This accounts for the minus sign in eqn. (\ref{zold}).}
\be\label{zold}I=-\frac{1}{4}\int_X\d^Dx\sqrt g \,\Tr \,F_{\mu\nu} F^{\mu\nu} \ee
and the field equations are
\be\label{wold} D^\mu F_{\mu\nu}=0. \ee  For future reference, we recall the identity
\be\label{yold} D^\mu(D^\nu F_{\mu\nu})=0,\ee
which reflects the fact that the Yang-Mills action is gauge-invariant even off-shell.

Now let $\A$ be a classical solution and set $A=\A+\a$.  However, it is clumsy to write $A_0$ all the time, so henceforth we will refer to the underlying
gauge field as $\h A$, and a chosen classical solution as $A$, and we will write the expansion as $\h A=A+\a$.  To linear order in $\a$, the classical equations become
\be\label{nold} D_\mu(D_\mu \a_\nu-D_\nu \a_\mu) +[F_{\nu\lambda},\a_\lambda]=0  \ee
where $D_\mu$ and $F_{\mu\nu}$ are the covariant derivative and field strength of the background solution $A$.
We can write this equation as $\Dw\a=0$ where 
\be\label{mold}(\Dw\a)_\nu=- D_\mu D^\mu\a_\nu+D_\nu D_\mu \a^\mu-2[F_{\nu\lambda},\a^\lambda]. \ee
The corresponding action, to quadratic order in $\a$, is
\be\label{pold} I'=-\frac{1}{2}\int_X\d^Dx\,\Tr\left( D_\mu\a_\nu D^\mu\a^\nu-(D_\mu \a^\mu)^2+2F_{\mu\nu}[\a^\mu,\a^\nu]\right). \ee

The operator $\Dw$ is not elliptic.  Its leading symbol is the matrix-valued function $\sigma(p)_{\mu\nu}=p^2\delta_{\mu\nu}-p_\mu p_\nu$
(tensored with the identity operator on the Lie algebra $\g$ of $G$),
and this matrix annihilates any vector that is a multiple of $p_\nu$.  This failure of ellipticity is an inevitable consequence of the
underlying 
gauge-invariance, which in terms of the linearization becomes 
\be\label{yoggo}\a_\mu\to \a_\mu-D_\mu\veps,\ee
for any $\g$-valued gauge parameter $\veps$. (We will abbreviate 
(\ref{yoggo}) as $\a\to \a-\d_A\veps$.)  The gauge-invariance implies that $\a_\mu=-D_\mu\veps$ is a solution
of $\Dw\a=0$ for any $\veps$, so $\Dw$ has an infinite-dimensional kernel and cannot possibly be elliptic.

To restore ellipticity, we need a suitable gauge condition.   A very natural one is $S=0$, where
\be\label{rold} S=D_\mu\a^\mu. \ee
When supplemented with the gauge condition $S=0$, the equation $\Dw\a=0$ becomes $\Dt\a=0$
where
\be\label{told} (\Dt\a)_\mu= -D_\nu D^\nu \a_\mu-2[F_{\mu\lambda},\a^\lambda]. \ee
The symbol of $\Dt$ is $p^2\delta_{\mu\nu}$, and this is invertible for nonzero real $p$, so $\Dt$ is elliptic. 
We note that
\be\label{sold}(\Dt\a)_\mu=(\Dw\a)_\mu-D_\mu S. \ee 

Now let us discuss why $S=0$ is a good gauge condition, first from the point of view of classical partial differential
equations and then in terms of quantum perturbation theory.

From the first point of view, we usually want to describe the solutions of the gauge-invariant equation
$\Dw\a=0$, modulo gauge transformations.   The claim is that such equations are in natural correspondence with
solutions of the gauge-fixed equation $\Dt\a=0$.   In one direction, if we are given an $\a$ that satisfies $\Dw\a=0$,
we look for a gauge-equivalent $\a'=\a-\d_A\veps$ with $S(\a')=0$.   The equation $S(\a')=0$ is equivalent to
\be\label{porom} P \veps= -S(\a),\ee
where $P =-D_\mu D^\mu$ is the gauge-invariant  Laplacian.   Eqn. (\ref{porom}) will have a unique
solution if the operator $P$ is invertible, and more generally if the right hand side is orthogonal to any zero-modes of $P$.
  We observe that
\be\label{yorom}-\int_X \d^Dx \sqrt g\sum_\mu\Tr\, (D_\mu\veps)^2=-\int_X \d^Dx\sqrt g \Tr \left(\veps P \veps\right). \ee
The left hand side is strictly positive unless $D_\mu\veps=0$, in which case it vanishes.  But if $\veps$ is an eigenfunction of $P$
with zero or negative eigenvalue, then the right hand side is zero or negative.
  This shows that $P$ is positive-definite -- and therefore invertible -- except for possible zero modes that must be
covariantly constant.  But $S(\a)=D_\mu\a^\mu$ is orthogonal to any covariantly constant mode, since
$\int_X \d^Dx\sqrt g \Tr\,\veps S(\a)=-\int_X\d^Dx\sqrt g\Tr\,D_\mu\veps \a^\mu$, which vanishes if $D_\mu\veps=0$.  The right hand
side of eqn. (\ref{porom}) is thus orthogonal to the kernel of $P$, so eqn. (\ref{porom}) always has a unique
solution for $\veps$.  

Thus, a solution of $\Dw\a=0$ is gauge-equivalent to a unique solution of the gauge-fixed equation $\Dt\a=0$.
In the opposite direction, we would like to prove that any solution of $\Dt\a=0$ actually obeys $\Dw\a=0$.
To show this, we first observe that the gauge-invariant operator $\Dw$ satisfies the identity
\be\label{orom} D_\mu((\Dw\a)^\mu)=0. \ee
This is proved by linearizing the underlying identity (\ref{yold}).  Comparing $\Dt$ and $\Dw$, it follows that
\be\label{forom}  D_\mu ((\Dt\a)^\mu)=-D_\mu D^\mu(D_\nu\a^\nu)=P S(\a). \ee
and therefore a solutions of $\Dt\a=0$ satisfies $P S(\a)=0$.
As we have just seen, the equation $P S=0$ implies that $S$ is covariantly constant,  $D_\mu S(\a)=0$.
But in view of eqn. (\ref{sold}),  $\Dt\a=0=D_\mu S(\a)$ implies that $\a$ satisfies the gauge-invariant
equation $\Dw\a=0$.    So it is equivalent to consider solutions of $\Dw\a=0$ up to gauge transformation or to
consider solutions of $\Dt\a=0$.  

For a fuller understanding, let us consider BRST quantization.  In BRST quantization, we introduce a ghost field $c$ that represents
a generator of gauge transformations, but with  fermionic statistics (and ghost number 1).  The BRST transformations of $\a$ and $c$ are 
\be\label{zofft}\delta \a_\mu=-D_\mu c ,~~~~\delta c=\frac{1}{2}[c,c]. \ee
   We also introduce an antighost multiplet consisting of an antighost field $\bar c$
and an auxiliary field $B$ in the adjoint representation, with
\be\label{wofft}\delta\bar c=B, ~~~~\delta B=0. \ee   
All this is consistent with $\delta^2=0$.
 The gauge-fixed action is obtained by adding
$\delta\int_X\d^Dx\sqrt g V$, for some convenient choice of $V$, to the gauge-invariant action (\ref{zold}).
Taking $V=\Tr\,(\frac{1}{2}\bar c B -\bar c D_\mu\a^\mu)$, we get 
\be\label{zoff}\delta\int_X\d^Dx\sqrt g V=\int_X\d^Dx\sqrt g \left(\frac{1}{2}B^2 -BD_\mu\a^\mu-\bar c D_\mu D^\mu c\right). \ee
Upon integrating out the auxiliary field $B$, we get the gauge-fixing action
\be\label{wiloff} I''=\int_X\d^Dx\sqrt g\Tr\left(-\frac{1}{2}(D_\mu\a^\mu)^2-\bar c D_\mu D^\mu c \right). \ee
The gauge-fixed action $\h I=I'+I''$ is 
\be\label{niloff}\h I = -\int_X\d^Dx\sqrt g\Tr\left(\frac{1}{2} D_\mu\a_\nu D^\mu\a^\nu+F^{\mu\nu}[\a_\mu,\a_\nu]+\bar c D_\mu D^\mu c \right). \ee
The kinetic operator for $\a$ is the gauge-fixed operator $\Dt$.

The fact that a BRST-invariant action can be written with $\Dt$
as the kinetic operator for gauge fields does {\it not} imply that $\Dt$ is elliptic.   Rather, a good BRST gauge-fixing -- suitable for perturbation theory -- is one
in which $V$ is chosen, as we have done in this example, to ensure
 that the resulting kinetic operators are elliptic.   Since ellipticity is an open condition, this means roughly that $V$ must be sufficiently generic.

In the above example, when we integrate out $B$, we get $B=S(\a)$, and therefore the BRST transformations become
\be\label{mork} \delta \bar c = S(\a). \ee

Of course, we do not have to integrate out $B$; we could develop a formalism with both $\a$ and $B$ present in the theory.
For the purposes of the present paper, this would lead to a somewhat more involved discussion, leading to the same conclusions.   That is because $\a$ obeys a second
order classical differential equation, while the equations of motion involve only first derivatives of $B$; thus we would be in the situation
described in footnote \ref{delfic} of section \ref{ellbc}.  To proceed in this way, we would  have to change the definition of the ``leading'' term in a differential equation by saying that $B$
has 
degree 1, just like a derivative $\partial/\partial x$. To avoid these complications, we will consider the theory with $B$ integrated out --
as is usually done in constructing Yang-Mills perturbation theory.   A similar remark applies later when we come to gravity.

\subsection{Yang-Mills Theory on a Manifold With Boundary}\label{ymbo}

Now let us suppose that $X$ has a boundary $M$, and try to extend this analysis to make a BRST-invariant
and elliptic gauge-fixing  in Yang-Mills theory in the presence of the boundary.  We assume that the gauge-fixed action is supposed
to be as above in the bulk, and we will discuss what we can do along the boundary.

We will try to implement a very natural boundary condition, in which one specifies the restriction of
the gauge field $A$ to the boundary.   We locally model $X$ by $x^D\geq 0$; 
  we denote the coordinates as $x^i$ for $i<D$ and we write $x^\perp$
for $x^D$.  Likewise we denote the gauge field components as $A_i$, $i<D$ and $A_\perp= A_D$.
For a boundary condition, we specify the boundary values of $A_i$, $i<D$ but not
of $A_\perp$.    In other words, we specify the gauge connection that would be used for parallel transport within $M$.
In terms of the field $\a$ that describes small fluctuations around a classical solution, this means that $\vec \a=
(\a_1,\a_2,\cdots, \a_{D-1})$ will vanish on $M$, but the condition on $\a_\perp=\a_D$ will be different.

If we want to impose a condition $\vec \a|=0$, then in view of the gauge-invariance $\a\to \a-\d_A\veps$,
we must constrain the generator $\veps$ of a gauge transformation to vanish along the boundary $M$.   
Since the ghost field $c$ is always the generator of a gauge transformation (with statistics reversed), it will also have
to vanish along the boundary.   Thus $c$ must obey Dirichlet boundary conditions:
\be\label{yorb} c|=0. \ee

Once we impose Dirichlet boundary conditions for $c$, we have to do the same for $\bar c$.  The reason is not that we want
to be able to interpret $\bar c$ as the complex conjugate of $c$, but that once we impose Dirichlet boundary conditions for $c$,
 we will not be able to define a sensible Green's function for the $c$-$\bar c$ system if we do anything else for $\bar c$.  
 Recall first of all that if we do impose Dirichlet boundary conditions on both $c$ and $\bar c$, then there is a standard Green's
 function 
$G(x,y)=\langle c(x)\bar c(y)\rangle$ that obeys the differential equations $\Delta_x G(x,y)=\Delta_yG(x,y)=\delta^D(x,y)$
(where $\Delta_x$ and $\Delta_y$ are  the gauge-invariant Laplacians  acting on the $x$  or $y$ variables)
along with the Dirichlet 
boundary conditions $G(x,y)|_{x\in M}=G(x,y)|_{y\in M}=0$.     But actually, just the equation $\Delta_x G(x,y)=\delta^D(x,y)$
plus the Dirichlet boundary condition in the $x$ variable 
$G(x,y)|_{x\in M} = 0$
uniquely determines $G(x,y)$.   The unique possibility is the standard Green's function
 that corresponds to imposing Dirichlet boundary conditions also on $\bar c$:
\be\label{lorb}\bar c|=0.\ee

We should not expect to be able to set $\a_D$ to zero along $M$, because $\a_D$ is not invariant under gauge
transformations whose generator $\veps$ vanishes on $M$.  To make $\a_D| $ invariant under $\a\to \a-\d_A\veps$,
we would want to require $D_\perp\veps|=0$.  This would entail restricting the 
generator  $\veps$ of a gauge transformation to satisfy $\veps|=D_\perp\veps|=0$, so 
we would want $c$ to satisfy $D_\perp c|=0$ as well as $c|=0$; but we cannot impose both Dirichlet and Neumann boundary conditions
on a field that obeys a second order wave equation.  

We can easily deduce from eqn. (\ref{lorb}) what boundary condition $\a_\perp$ must satisfy.   Given that $\bar c$ vanishes along
$M$ and given the BRST transformation law $\delta \bar c=S(\a)$ (eqn. (\ref{mork})), BRST invariance requires
\be\label{ocor}S(\a)|=0.\ee
If $X$ is the half-space $x_\perp\geq 0$ in $\R^D$, then given our condition $\vec \a|=0$,   $S(\a)|$ reduces to $D_\perp \a_\perp|$
and thus $\a_\perp$ satisfies a gauge-invariant version of Neumann boundary conditions:
\be\label{cor}D_\perp \a_\perp|=0.  \ee
In the general case of a curved manifold, $S(\a)|=0$ corresponds to a boundary condition on $\a_\perp$
 that is similar locally to Neumann boundary conditions on a scalar field,
plus a lower order term.  We will loosely refer to this as Neumann boundary conditions for $\a_\perp$.

The fact that the leading symbol of $\Dt$ is $p^2\delta_{\mu\nu}$ means that it behaves
at short distances as a system of decoupled scalar Laplace equations
for the components $\a_1,\a_2,\cdots,\a_D$.   Dirichlet and Neumann boundary conditions on the scalar Laplace equation are elliptic.
So the  combination of Dirichlet boundary conditions for $\vec\a$ and Neumann for $\a_\perp$ comprises an elliptic boundary condition for the
operator $\Dt$, and thus we have found a BRST-invariant and elliptic gauge-fixing for Yang-Mills theory on
a manifold with boundary.

Now let us discuss how one might motivate this boundary condition from the point of view of differential geometry, without mentioning
the ghosts.  In that framework, one may want to study solutions of the gauge-invariant equation $\Dw\a=0$, with $\vec \a|=0$,
modulo gauge transformations that are trivial along the boundary.  One wishes to show that such solutions are in one-to-one correspondence
with solutions of the gauge-fixed equation $\Dt\a=0$, with boundary conditions
$\vec \a|=S(\a)|=0$.   In one direction, given a solution of $\Dw\a=0$,
we look as in section \ref{ym} for a gauge transformation $\a\to\a-\d_A\varepsilon$ that will  set $\Dt\a=0$.  Given the relation between
$\Dt$ and $\Dw$ (eqn. (\ref{sold})), this means that we want to set $D_\mu S(\a)=0$.  Since we also want to satisfy $S(\a)|=0$, we actually need $S(\a)$ to vanish identically.  So we have to find a gauge parameter $\varepsilon$ such
that $\varepsilon|=0$ and $P\varepsilon=-S(\a)$.   Essentially the same argument as in section \ref{ym} shows that with Dirichlet
boundary conditions, $P$ is invertible.  So there is a 
unique solution of $P\varepsilon=-S(\a)$.

In the opposite direction, if we are given a solution of $\Dt\a=0$ that also satisfies the boundary condition $S(\a)|=0$, we want to show
that actually $S(\a)$ vanishes identically, so that the gauge-invariant equation $\Dw\a=0$ is satisfied.   As before, the  identity
(\ref{forom}), together with $\Dt\a=0$, implies that $P S(\a)=0$, and this, together with the boundary condition $S(\a)|=0$, implies
that $S(\a)$ vanishes identically.

\section{Elliptic Boundary Conditions In Gravity}\label{three}

In discussing elliptic gauge-fixing and elliptic boundary conditions in gravity, we will be brief on points on which there is a very close parallel with what we have
already described for Yang-Mills theory.

\subsection{General Relativity on A Closed Manifold}\label{grclosed}

\def\hh{{\sf h}}
\def\veps{\varepsilon}
The action of classical General Relativity with a cosmological constant $\Lambda$   is
\be\label{gract} I = -\frac{1}{\kappa^2} \int_M \d^Dx \sqrt{  g}\left(R - 2\Lambda\right). \ee
We can of course add matter fields, but we do not do so explicitly.
The field equations read
\be\label{ract}R_{\mu\nu}-\frac{1}{2}g_{\mu\nu}R+\Lambda g_{\mu\nu}=0,\ee
and are governed by a Bianchi identity
\be\label{wact}D^\mu(R_{\mu\nu}-\frac{1}{2}g_{\mu\nu}R+\Lambda g_{\mu\nu})=0. \ee
We expand around a background classical solution $g_0$ with $g_{\mu\nu}=g_{0\mu\nu}+h_{\mu\nu}$.   However, again it is clumsy to always write
$g_{0\mu\nu}$ for the background field, so we will instead write $\h g_{\mu\nu}$ for the full metric, and $g_{\mu\nu}$ for the background, so that the
expansion reads $\h g_{\mu\nu}=g_{\mu\nu}+h_{\mu\nu}$.  
Covariant derivatives and curvatures
will refer to the background metric, which is also used in raising and lowering indices. It is convenient to define $\hh=h^\mu_\mu=g^{\mu\nu}h_{\mu\nu}$.

 The quadratic part of the action for $h$ is, of course, an important input in semiclassical quantization \cite{GP,CD}.
The $D$-dimensional version of the formula, from for example eqn. (2.3) of \cite{BB}, is\footnote{In the absence of matter fields, the following could be
simplified slightly using the equations of motion for the background field to replace $R_{\mu\nu}$ with a multiple of $g_{\mu\nu}$.  The reason that we do not do that is that we wish to make statements that
remain valid if matter fields are included. (Admittedly, we are following a hybrid logic, since we do not add explicitly the matter contributions
to the action or the equations of motion.) In eqn. (\ref{quadact}), the terms proportional to $\Lambda$ are those that come from the $\Lambda$ term in the original
action (\ref{gract}).}
\begin{align}\label{quadact} I'=-\frac{1}{\kappa^2}\int_X\d^D x \sqrt g& \left(\frac{1}{4}h^{\mu\nu} (D_\lambda D^\lambda +2\Lambda)h_{\mu\nu}-\frac{1}{8}\hh(D_\lambda D^\lambda+2\Lambda)\hh+\frac{1}{2}(D^\nu h_{\mu\nu}-\frac{1}{2}
\partial_\mu\hh)^2\right.\cr & \left.+\frac{1}{2}h^{\mu\lambda}h^{\nu\rho}R_{\mu\nu\lambda\rho}+\frac{1}{2}\left(h^{\mu\lambda}h^\nu_\lambda-\hh h^{\mu\nu}\right)R_{\mu\nu}+\frac{1}{8}(\hh^2-2h^{\mu\nu}h_{\mu\nu})R\right).\end{align}
The gauge-invariant linear wave equation satisfied by $h$ is $(\D h)_{\mu\nu}=0$ where we set $2\kappa^2\delta I'/\delta h^{\mu\nu}=(\D h)_{\mu\nu}$.
We will not write explicitly the rather unilluminating formula for $\D$.  Gauge-invariance implies of course that the operator $\D$ is not elliptic, but it satisfies a Bianchi identity that descends directly from the underlying
Bianchi identity (\ref{wact}):
\be\label{newb}D^\mu((\D h)_{\mu\nu})=0.\ee 

To restore ellipticity and carry out quantum perturbation theory, we need a gauge condition.  The form of the action suggests a convenient and widely used
choice of gauge (variously known as harmonic, de Donder, or Bianchi gauge), namely $T_\mu(h)=0$ where
\be\label{yelf} T_\mu(h)= D^\nu h_{\mu\nu}-\frac{1}{2} \partial_\mu \hh. \ee
To decide if this is a good gauge condition, we have to ask if it can be implemented by an infinitesimal coordinate transformation 
\be\label{infco} \delta h_{\mu\nu}=D_\mu\veps_\nu+D_\nu\veps_\mu.\ee
We observe that 
\be\label{melf}\delta T_\mu(h)= D_\nu D^\nu\veps_\mu+R_{\mu\nu}\veps^\nu. \ee 
So to set $T_\mu(h)=0$, we need to solve 
\be\label{lelf} (P\veps)_\mu = -T_\mu(h),\ee
where the operator $P$ is defined by \be\label{hurz}(P\veps)_\mu =-D_\nu D^\nu\veps_\mu -R_{\mu\nu}\veps^\nu.\ee

  We note that this operator is elliptic and thus has a discrete spectrum.
  If $P$  is invertible, there will be a unique solution $\veps$ of eqn. (\ref{lelf}) and thus
the gauge condition is good.  
The operator $P$ is invertible in pure gravity with $\Lambda<0$.  This makes $R_{\mu\nu}$ negative-definite, and since the Laplace-like
operator
$-D^\mu D_\mu  $ is positive semi-definite (by the same argument as in eqn. (\ref{yorom})),  $P$ is then strictly positive.  Even if the cosmological constant is not negative or matter fields are present, 
one can reasonably expect that
in expanding around a generic classical solution, $P$ will have no zero-mode.\footnote{With positive cosmological constant, for $X=S^4$, $P$ has zero and negative
eigenvalues that correspond to Killing vectors and conformal Killing vectors \cite{CD}.}  However, if $P$ does have a zero-mode, then the gauge fixing procedure needs to
be slightly modified to treat this mode correctly. Being elliptic, $P$ will never have more than finitely-many zero-modes on a compact manifold.   It is technically inconvenient
to have to slightly modify the gauge-fixing condition, but as this only affects finitely many modes, it does not really affect any questions of principle.
Actually, on a manifold with nonempty boundary, which is our main interest in this paper, this complication does not arise, in the following sense.  It is shown in Lemma 2.2
of \cite{Anderson} that, acting on vector fields that are required to vanish on the boundary, the operator $P$ is always invertible, regardless of $\Lambda$.  (A key step
in the proof is the fact that a Killing vector field that vanishes along the boundary is identically zero.)

To implement the gauge condition $T_\mu=0$ in the BRST framework, we first add the ghosts, which are a fermion field $c^\mu$ that represents the generator of a diffeomorphism
(that is, $c^\mu$ transforms as a vector field), with BRST variations
\be\label{helf} \delta h_{\mu\nu}=D_\mu c_\nu+D_\nu c_\mu, ~~~~~ \delta c^\mu = c^\nu\partial_\nu c^\mu. \ee
One also needs an antighost multiplet consisting of the antighost field $\bar c^\mu$ and an auxiliary field $f^\mu$, with
\be\label{elf}\delta \bar c^\mu=f^\mu,~~~~\delta f^\mu=0.  \ee 
A convenient gauge-fixing term is
\be\label{welf}\frac{1}{\kappa^2}\delta\int_X\d^Dx \sqrt g\left(-\frac{1}{2}\bar c_\mu f^\mu +\bar c^\mu T_\mu(h)\right). \ee 
After integrating out the auxiliary field, this generates a correction to the gravitational action
\be\label{nelf}I''=\frac{1}{2\kappa^2}\int_X\d^Dx \sqrt g \,T_\mu(h)^2=\frac{1}{2\kappa^2}\int_X\d^Dx \sqrt g \left(D^\nu h_{\mu\nu}-\frac{1}{2}
\partial_\mu\hh\right)^2.\ee
The gauge-fixed gravitational action is
\begin{align}\label{dolf} I'+I''= -\frac{1}{\kappa^2}\int_X&\d^D x \sqrt g \left(\frac{1}{4}h^{\mu\nu} (D_\lambda D^\lambda+2\Lambda)h_{\mu\nu}-\frac{1}{8}\hh(D_\lambda D^\lambda+2\Lambda)\hh+\right.\cr & \left.+\frac{1}{2}h^{\mu\nu}h^{\rho\sigma}R_{\mu\nu\rho\sigma}+\frac{1}{2}\left(h^{\mu\lambda}h^\nu_\lambda-\hh h^{\mu\nu}\right)R_{\mu\nu}+\frac{1}{8}(\hh^2-2h^{\mu\nu}h_{\mu\nu})R\right).   \end{align}
The action for the ghosts is
\be\label{igh} I_{\mathrm{gh}}=\frac{1}{\kappa^2}\int_X\d^Dx\sqrt g \,\bar c P c = \frac{1}{\kappa^2}\int_X\bar c^\mu\left(-g_{\mu\nu} D_\lambda D^\lambda -R_{\mu\nu}\right) c^\nu. \ee   The gauge-fixed linear kinetic operator $\D'$ that governs metric fluctuations is defined by
$2\kappa^2\delta (I'+I'')/\delta h^{\mu\nu}=(\D' h)_{\mu\nu}$
From eqn. (\ref{dolf}), we see that the leading symbol $\sigma_x(p)$ of $\D'$ is invertible,
but not positive-definite. (This lack of positivity was first pointed out and discussed in \cite{GHP}.) In fact, $\sigma_x(p)$  acts on the traceless part of $h_{\mu\nu}$ as
 a positive multiple of $p^2$ (tensored with the identity matrix on the $\mu\nu$ indices) and on the trace $\hh$ as a negative multiple of $p^2$.    Thus $\D'$ is elliptic,
 even though not positive-definite.

Actually, in the absence of matter fields, the traceless and trace parts of $h_{\mu\nu}$ decouple (as one sees by using the classical equations of
 motion to replace the background $R_{\mu\nu}$ with a multiple of $g_{\mu\nu}$) and can be treated separately in constructing propagators and determinants.  In this
 case, the eigenvalues of $\D'$ are almost all positive on traceless modes and almost all negative on trace modes.  
 In the presence of matter fields, the traceless and trace modes do not decouple in general, but ellipticity ensures that $\D'$ still has a discrete spectrum.
 Large positive eigenvalues correspond to wavefunctions that are almost traceless, and large negative eigenvalues correspond to wavefunctions whose traceless part is
 very small.

 Ellipticity guarantees that  $\D'$ has at most finitely many zero-modes.  As usual, these modes must be treated specially in constructing perturbation theory.
 Ellipticity guarantees that $\D'$ always has a parametrix, that is, a propagator suitable for perturbation theory.

 To construct perturbation theory, in addition to a propagator and a renormalization
 procedure, one requires a one-loop determinant, since  the one-loop path integral formally includes a factor $\det P/\sqrt{\det \D'}$.    Here as the operators in question
 are elliptic, $\zeta$-function regularization can be straightforwardly used to define the absolute values of the determinants, but the presence of negative eigenvalues
 -- infinitely many of them in the case of $\D'$ -- means that it is not straightforward to understand the phase of the one-loop path integral.   This issue has been
 discussed in several papers \cite{GHP,Polch}, but its status is not entirely clear.
 We will not discuss these questions here except to note that existing computations rely on the decoupling of the traceless and trace parts of the metric
 that holds in pure gravity, so at a minimum some generalization is needed.
 
 From the form (\ref{nelf}) of the gauge-fixing part of the gravitational action, one can work out an explicit formula relating $\D'$ and $\D$:
 \be\label{worm} (\D' h)_{\mu\nu}=(\D h)_{\mu\nu}+D_\mu T_\nu(h)+D_\nu T_\mu(h)- g_{\mu\nu} D_\lambda T^\lambda(h).\ee
 From this and the Bianchi identity (\ref{newb}), one finds that
 \be\label{prom} D^\mu((\D' h)_{\mu\nu}) = (P T(h))_\nu. \ee
 Thus the equations of motion of the gauge-fixed theory imply
 \be\label{rom} (PT(h))_\mu=0.\ee
 This statement remains valid when matter fields are included (assuming that the gauge-fixing takes the form of eqn. (\ref{nelf})),  since the Einstein equations
 with matter fields included still satisfy a Bianchi identity, which reflects the underlying general covariance.  (The proof of the general Bianchi identity requires the equations
 of motion for the matter fields as well as the metric.)

 From the point of view of BRST quantization, the procedure that we have described is satisfactory if $\D'$ and $P$ have no zero-modes, as one may expect
in expanding around a generic classical solution.    In general, the procedure needs to be slightly modified to treat properly a finite-dimensional space of zero-modes.

 Let us now discuss how would one motivate this procedure from the point of view of differential geometry, without reference to quantization.
  From that point of view, one would like to compare the solutions of the gauge-invariant equation $\D h =0$, modulo the gauge equivalence (\ref{infco}), to the solutions of the
 gauge-fixed equation $\D' h = 0$.  In one direction, we have already seen that if $P$ is invertible, then every solution of $\D h=0$ can be uniquely put in a gauge
 with $T_\mu(h)=0$.  Then eqn. (\ref{worm}) shows that $\D' h = 0$.  In the opposite direction, if $\D' h=0$, then from eqn. (\ref{prom}), we have $(PT(h))_\mu=0$, which
 (if $P$ is invertible) implies that $T_\mu(h)=0$.   Then using eqn. (\ref{worm}) again, we see that the gauge-invariant equation $\D h=0$ is satisfied.

\subsection{General Relativity on A Manifold With Boundary}\label{grbound}

We now consider General Relativity on a manifold $X$ with boundary $M$.  We start by analyzing the most direct analog of the boundary condition for Yang-Mills
theory that was discussed in section \ref{ymbo}.   

In this boundary condition, we keep fixed the boundary metric of $M$ and allow fluctuations in the interior.   Thus if $X$ is locally defined by $x^D\geq 0$, while $M$ is parametrized by\footnote{As before, we will write $\vec x$ and $x_\perp$ for tangential coordinates
$x^1,\dots, x^{D-1}$ and the normal coordinate $x^D$.} $x^i$, $i=1,\cdots, D-1$,   we specify the boundary values of $g_{ij}$.   In terms of the metric
perturbation $h_{\mu\nu}$, this means that part of the boundary condition will be 
\be\label{murky} h_{ij}|=0, ~~~~~~~i,j=1,\cdots ,D-1. \ee
The boundary conditions on $h_{i\perp}$ and $h_{\perp\perp}$ are still to be specified.

To learn what the remaining boundary conditions will have to be, we first consider the gauge symmetries $\delta h_{\mu\nu}=D_\mu\veps_\nu+D_\nu\veps_\mu$.
If $\veps$, restricted to $M=\partial X$, has a nonzero component in the normal direction, then it does not really generate a symmetry of $X$, as it tries to move
the boundary of $X$ normal to itself.   Thus the diffeomorphism group of $X$ is generated by vector fields that are constrained
by
\be\label{ohat}\veps^\perp|=0. \ee
In addition, if we wish to impose a boundary condition $h_{ij}|=0$, we must restrict ourselves to vector fields with 
\be\label{nohat} \veps^i|=0.\ee 
Combining these two statements, we see that we should consider only diffeomorphisms generated by vector fields that satisfy
\be\label{phat}\veps^\mu|=0. \ee

In BRST quantization, this means that the ghost field $c^\mu$ should satisfy Dirichlet boundary conditions
\be\label{rhat} c^\mu|=0.\ee
Now let us assume that the gauge-fixing away from the boundary is carried out by the procedure of section \ref{gract}.  
For the same reason as in section \ref{ymbo}, the antighost field $\bar c^\mu$ must likewise satisfy Dirichlet boundary
conditions:
\be\label{dhat} \bar c^\mu|=0. \ee
On the other hand, after eliminating the auxiliary field,
the BRST variation of  $\bar c^\mu$ is
\be\label{mhat}\delta \bar c^\mu = T^\mu(h), \ee
which is the direct analog of eqn. (\ref{mork}) for Yang-Mills theory.  
Therefore, BRST invariance forces us to impose
\be\label{lhat} T_\mu(h)|=0, \ee
similarly to eqn. (\ref{ocor}) in gauge theory.   

Eqn. (\ref{lhat}) is a boundary condition for $h_{\perp\perp}$ and $h_{i\perp}$, somewhat analogous to Neumann boundary conditions. 
Together with eqn. (\ref{murky}), it gives the right number of conditions to make a boundary condition for metric fluctuations.   For brevity we will
call this the Dirichlet boundary condition.
However \cite{Anderson,AE}, this boundary condition is not elliptic.
We will first show this by a short computation and then give a less computational explanation.

 Since the considerations are local and only depend on the leading symbol of the linearized
Einstein equations and the leading behavior of the boundary condition at short distances, we can take $X$ to be a half-space $\R^D_+$
in a flat Euclidean space $\R^D$, say the half-space $x_\perp\geq 0$. 
 A general plane wave solution with
nonzero momentum $\vec k$ along the boundary that decays exponentially for large $x_\perp$
takes the form
\be\label{znolt}h_{\mu\nu}=\alpha_{\mu\nu} e^{\i \vec k\cdot \vec x-|\vec k| x_\perp}. \ee
To show that the boundary condition is not elliptic, we have to show that it is possible for a solution of this kind with real nonzero $\vec k$ 
to satisfy the boundary conditions.   (Since the boundary conditions are invariant under scaling of $\vec k$, if we can satisfy them for any nonzero $\vec k$
we can do so with arbitrarily large $|\vec k|$.)   Dirichlet boundary conditions $h_{ij}|=0$ imply that we should set $\alpha_{ij}=0$ in eqn. (\ref{znolt}).
Let us  write $\vec\alpha$ for the $(D-1)$-vector
with components $\alpha_{i\perp}$, and $\beta$ for $\alpha_{\perp\perp}$. 
A short computation reveals that the equations $T_{\perp\perp}=0$ and $T_{i\perp}=0$ become
\be\label{mollt} \i \vec k\cdot \vec \alpha-\frac{1}{2}|\vec k|\beta=0\ee
and
\be\label{zolott} -|\vec k|\vec\alpha -\frac{\i}{2}\vec k \,\beta=0. \ee 
We can satisfy both of these equations with
\be\label{polt}\vec \alpha=-\frac{\i}{2|\vec k|}\vec k \beta,\ee
and therefore Dirichlet boundary conditions for gravity are not elliptic. 

At first this may look like an unlucky accident, and one may wonder if using a different bulk gauge condition would have
avoided the problem.   This is not the case, as is shown in several ways in  \cite{Anderson}.   One  argument makes use of the second order
nature of the Hamiltonian constraint equation of General Relativity.
A second argument is as follows.   We will describe a compact $X$ with boundary such that the linearized Einstein equations on $X$, without
any gauge-fixing, and 
with Dirichlet boundary conditions $h_{ij}|=0$ (but no boundary condition placed on $h_{i\perp}$ or $h_{\perp\perp}$), has infinitely many zero-modes, modulo gauge transformations.   Any correctly gauge-fixed
version of the linearized Einstein equations on $X$ would have the same infinite-dimensional kernel, contradicting ellipticity.  

We take $X$ to be a product $T\times I$, where $T$ is a torus with flat metric, parametrized by periodic variables $\vec x=(x_1,\cdots, x_{D-1})$,
 and $I$ is the unit interval $0\leq x_\perp\leq 1$.  Now we pick a function $\veps(\vec x)$ and perturb $X$ so that its boundaries are
 $\veps(\vec x)\leq x_\perp \leq 1$.  Since the extrinsic curvature of $\partial X$ vanishes, the boundary geometry of $X$ is unchanged to first order in $\veps$,
 and therefore these perturbations satisfy $h_{ij}|=0$.    On the other hand, these perturbations cannot be eliminated by a diffeomorphism.
 So with Dirichlet boundary conditions,
  the kernel of the linearized Einstein equations on $X$, modulo its subspace induced by diffeomorphisms of $X$, is infinite-dimensional.   Even though this is a very
  special example, it is enough to show that linearized Einstein equations with Dirichlet boundary conditions, and with any choice of gauge-fixing, cannot be elliptic.

It is instructive to see how to put the space of zero-modes that we found in this example in the form of a change
in  the metric ($g\to g+h$) rather than a change in the range of the coordinates.   Since the perturbation preserves the flatness of $X$,
it must take the form
\be\label{yrd}h_{\mu\nu}=\partial_\mu v_\nu +\partial_\nu v_\mu\ee
for some vector field $v^\mu(\vec x,x_\perp)$.  However, $v^\mu$ will not vanish on the boundaries of $X$.  Rather, we take $v^\mu$ to vanish at $x_\perp=1$,
but at $x_\perp=0$, we impose 
\be\label{wyrd} \vec v = 0,  ~~~~ v_\perp =\veps(\vec x). \ee
(In other words, $v^\mu\partial/\partial x^\mu|_{x_\perp=0} =\veps(\vec x)\partial/\partial x_\perp$.)
There is a unique $v^\mu$ that satisfies these boundary conditions and also satisfies 
\be\label{nyrd} P v = 0 .\ee
This condition ensures  that $h_{\mu\nu}$, defined in eqn. (\ref{yrd}), obeys $T_\mu(h)=0$.   With such a 
 choice of $v$,  the perturbations (\ref{yrd}) are nontrivial zero-modes of the linearized
Einstein equations on $X$ in the gauge $T_\mu(h)=0$.  Of course, the connection between the two descriptions is that a diffeomorphism generated by $v^\mu$,
to first order, maps the interval $0\leq x_\perp\leq 1$ to $\veps(\vec x)\leq x_\perp \leq 1$.   In particular,  because $v_\perp\not=0$ at $x_\perp=0$, $v$ does not generate a diffeomorphism of $X$.

\def\la{\langle}\def\ra{\rangle}\def\t{\widetilde}
We will discuss one last topic before moving on to the conformal boundary condition.
Assuming as above that the bulk gauge-fixing is carried out by adding $g^{\mu\nu}T_\mu(h)T_\nu(h)$ to the action, the boundary condition $T_\mu(h)|=0$ has another
virtue that we have not yet explained: it is needed to make the gauge-fixed linearized Einstein operator $\D'$ hermitian, in the following sense.
Let $\la h, \t h\ra$ be the obvious inner product on the space of metric deformations,
\be\label{pelg}\la h,\t h\ra=\int_X\d^Dx\sqrt g g^{\mu\nu}g^{\mu'\nu'} h_{\mu\mu'}\t h_{\nu\nu'}. \ee
Then $\D'$ is hermitian in the sense that
\be\label{welg} \la h,\D'\t h\ra=\la \D' h,\t h\ra. \ee
When one tries to prove this by integration by parts, one runs into surface terms.  However, the surface terms cancel with the help of the boundary conditions
that we have assumed.

It is not hard to prove this by hand, but a better explanation is as follows.
First of all, the gauge-invariant linearized Einstein operator $\D$ satisfies the same identity
\be\label{pelog}\la h,\D\t h\ra =\la \D h,\t h\ra.\ee
The most natural way to prove this is to use the fact that either the left or the right hand side can be interpreted as the quadratic part of the 
action, expanded around a chosen classical solution.   
However, for this to be  true, one must add a boundary term to the action.  The boundary term is chosen to ensure that, when one varies
the action to derive the equations of motion, the boundary terms in the variation of the action vanish, once the boundary conditions are imposed.   
Of course, the boundary term that will make this work, if there is one, depends on what boundary condition one wants.
With Dirichlet boundary conditions, the relevant boundary term is the Gibbons-Hawking-York GHY term \cite{GH,York};  with the
conformal boundary condition, a slightly different boundary term is appropriate \cite{Anderson3}, as we will explain in section \ref{works}. 
The variation of the Einstein-Hilbert action
\be\label{gracto} I = -\frac{1}{\kappa^2} \int_M \d^Dx \sqrt{  g}\left(R - 2\Lambda\right)\ee under $\delta g_{\mu\nu}=h_{\mu\nu}$
has the usual bulk term related to Einstein's equations and also a boundary term\footnote{In the following, $g_\partial=g|$ is the induced metric of the boundary
and so $\d^{D-1}x\sqrt{g_\partial}$ is the natural Riemannian measure of the boundary.}
\be\label{pacto}\delta_{\mathrm{bdry}}I=-\frac{1}{\kappa^2}\int_{\partial M}\d^{D-1}x \sqrt{g_\partial}\left(-2\delta \K-K^{ij}h_{ij}\right). \ee
Here $K_{ij}$ is the extrinsic curvature of the boundary and $\K=g^{ij}K_{ij}$ is its trace.   $\delta \K$ is the variation of $\K$ under $\delta g_{\mu\nu}=h_{\mu\nu}$.
We do not need the explicit formula for $\delta\K$, since this contribution to  the variation of the Einstein-Hilbert action is canceled by adding the GHY term
\be\label{racof}I_{{\mathrm{GHY}}}=-\frac{2}{\kappa^2} \int_{\partial M}\d^{D-1}x\sqrt{g_\partial} \,\K. \ee
 Dirichlet boundary conditions $h_{ij}|=0$ ensure the vanishing of the remaining term $K^{ij}h_{ij}$ in $\delta_{\mathrm{bdry}}I$ and the absence of any
  contribution from varying
$\sqrt {g_\partial}$ in $I_{{\mathrm{GHY}}}$.  So  with Dirichlet boundary conditions, there is no boundary term in the variation of the combined action $I+I_{{\mathrm{GHY}}}$.
This also means that there is no boundary term in proving (\ref{pelog}).  

In gauge fixing, we added to the gravitational action another term
\be\label{neelf}I''=\frac{1}{2\kappa^2}\int_X\d^Dx \sqrt g \,g^{\mu\nu} T_\mu(h)T_\nu(h).\ee
When we vary this under $\delta g_{\mu\nu}=h_{\mu\nu}$, upon integrating by parts to derive
 the bulk equations of motion, we generate additional surface terms.  But because $I''$ is bilinear in $T_\mu(h)$, these new
surface terms  are all proportional to $T_\mu(h)$ and so vanish if the boundary condition includes $T_\mu(h)=0$.  
With this being so, the left and right hand sides of eqn. (\ref{welg}) are both equal to the gauge-fixed quadratic action, so in particular they are equal.

\subsection{A Boundary Condition That Works}\label{works}

Though the Dirichlet boundary condition is not elliptic, there is a simple elliptic boundary condition for Einstein's equations \cite{Anderson}.    Instead of specifying the boundary metric, we specify
only the conformal structure of the boundary.   Differently put, we specify the boundary metric $\h g_{ij}|$  only up to a Weyl transformation $\h g_{ij}\to e^\phi \h g_{ij}$.
We write $\ovg$ for the conformal structure of the boundary, that is, for the equivalence class of the boundary metric, modulo a Weyl transformation.

In terms of the expansion $\h g_{\mu \nu}=g_{\mu\nu}+h_{\mu\nu}$, specifying only the conformal structure of the boundary
means that only the traceless part of the perturbation $h_{ij}|$ of the boundary metric 
is required to vanish, so that
\be\label{zolt} h_{ij}| = g_{ij} \gamma \ee
for some function $\gamma$.

We assume that the bulk gauge-fixing is that of section \ref{grclosed} and therefore, as in section \ref{grbound}, 
part of the boundary condition will be
\be\label{olt} T_\mu(h)|=0. \ee
We need one more boundary condition, to compensate for relaxing the constraint on the trace of $h_{ij}|$.   For this, we impose a constraint
on the trace of the extrinsic curvature.   We will write $K_{ij}$ for the extrinsic curvature in the metric $g$, and $\h K_{ij}$ for the extrinsic curvature
in the metric $\h g=g+h$.  We also 
 write $\h\K=\h g^{ij} K_{ij}(\h g)$ for the trace of the extrinsic curvature in the full metric $\h g=g+h$, and similarly $\K=g^{ij}K_{ij}(g)$ for the trace of
the extrinsic curvature computed using the background metric $g$.  
Then we complete the boundary condition by requiring 
\be\label{molt} \h \K=\K. \ee  In other words, the condition is that the perturbation does not change the trace of the extrinsic curvature.  
We call the combination of eqns. (\ref{zolt})-(\ref{molt}) the conformal boundary condition.

Explicitly, the linearization of eqn. (\ref{molt}), in coordinates in which the background metric satisfies $g_{\perp\perp}=1$, $g_{\perp i}=0$, is
\be\label{folt} D_\perp h^i{}_i -2 D^i h_{\perp i}+2h^{ij}K_{ij}=0. \ee
The term $2h^{ij}K_{ij}$, being of lower order, does not affect the discussion of ellipticity.

To show that the conformal boundary condition is elliptic, it suffices again to take $X=\R^D_+$ and to analyze 
  solutions of the equation $\D' h=0$ that propagate  like a plane wave along the boundary:
 \be\label{nolt}h_{\mu\nu}=\alpha_{\mu\nu} e^{\i \vec k\cdot \vec x-|\vec k| x_\perp}. \ee
We have to show that for any nonzero real $\vec k$, a solution of this kind can satisfy the boundary condition only if
$\alpha_{\mu\nu}=0$.

As a first step, we see that eqn. (\ref{zolt}) implies that $\alpha_{ij}=\delta_{ij} \gamma$ for some $\gamma$.   As before, we  write $\vec\alpha$ for the $(D-1)$-vector
with components $\alpha_{i\perp}$, and $\beta$ for $\alpha_{\perp\perp}$.   In this geometry, eqn. (\ref{folt}) reduces to
\be\label{polot} 2\i \vec k\cdot \vec \alpha +|\vec k|(D-1) \gamma = 0. \ee
The equations $T_0(h)|=0$ and $T_i(h)|=0$  become 
\be\label{wolt} \i \vec k\cdot \vec \alpha-\frac{1}{2}|\vec k|\beta+|\vec k|\frac{D-1}{2} \gamma =0\ee
and
\be\label{noltt} -|\vec k|\vec\alpha -\frac{\i}{2}\vec k \,\beta-\frac{\i}{2}\vec k (D-3)\gamma=0. \ee 
For nonzero real $\vec k$, these equations imply that $\vec \alpha=\beta=\gamma=0$, so the conformal boundary condition is elliptic.

The gauge-fixed linearized Einstein operator $\D'$  with conformal boundary conditions is not just elliptic but self-adjoint.  
This can be proved by modifying the discussion at the end of section \ref{grbound}.   With conformal boundary conditions,
a different normalization is needed for the GHY boundary term in the action \cite{Anderson3}.  The boundary variation $\delta_{\mathrm{bdry}}I$ of
the Einstein-Hilbert action is given by eqn. (\ref{pacto}) irrespective of the boundary conditions.  However, with the conformal
boundary condition, $\delta\K=0$ (since the conformal boundary condition is defined by keeping $\K$ fixed) but we no longer have  $h_{ij}|=0$; instead $h_{ij}|=g_{ij}\gamma$ for some scalar function $\gamma$.
This means that $\delta_{\mathrm{bdry} }I$ is now equal to $(1/\kappa^2)\int_{\partial M}\d^{D-1}x\sqrt {g_\partial}\, \K\gamma$.  To cancel this,
for a conformal boundary we need a boundary term $I_{\mathrm{CB}}$ that is a multiple of the usual GHY term:
\be\label{belz} I_{\mathrm{CB} }= \frac{1}{D-1}I_{{\mathrm{GHY}}}= -\frac{2}{D-1}\frac{1}{\kappa^2}\int_{\partial M}\d^{D-1}x\sqrt{ g_\partial}\,\K. \ee
The identity $\la h,\D \t h\ra=\la \D h,\t h\ra$ now holds, just as before, because the left and right hand sides are both equal to the quadratic
action derived from $I+I_{\mathrm{CB}}$.   And likewise, after adding the usual bulk gauge-fixing term to the action and imposing
the boundary condition $T_\mu(h)|=0$, the gauge-fixed operator $ \D'$ obeys the same identity.

  Self-adjointness gives a natural
identification between the kernel and cokernel of $\D'$, which in particular have the same dimension, generically zero.   Self-adjointness
 also means that the absolute value of the one-loop determinant
can be straightforwardly defined using zeta-function regularization.  (As remarked in section \ref{grclosed}, the phase of the determinant is more subtle.)

\subsection{Expanding or Contracting Metrics}\label{plumb}

In General Relativity, the metric on an initial value surface and the extrinsic curvature are canonically conjugate variables.  We seem to have learned that
at least  in Euclidean signature,
it is better to fix the conformal structure of the boundary and the trace of the extrinsic curvature, rather
than 
constraining all of the boundary metric.  This suggests that in quantization, one should consider a wavefunction $\h\Psi(\ovg,\K)$ that depends on 
the conformal structure of a hypersurface and the trace of the extrinsic curvature, rather than a wavefunction $\Psi(g)$ 
that depends on the metric of the hypersurface. (See \cite{York} for early ideas along these lines.)   There is, however, a further important detail that may change the picture, at least for many applications.
Even though Dirichlet boundary conditions are not elliptic, some of the important consequences of ellipticity do hold for Dirichlet boundary conditions, for a fairly wide class of metrics.\footnote{See section 3 of \cite{Anderson2}  for the  boundary condition that we are about to describe and its properties.
For antecedents of some of the ideas in a different context, see \cite{Hamilton}, pp. 187-93.}  

Let us replace eqn. (\ref{zolt}) with
\be\label{exc} h_{ij}|= K_{ij} \gamma    ,\ee
with an unspecified function $\gamma(\vec x)$, where again $K_{ij}$ is the extrinsic curvature of the background metric.  We leave eqns. (\ref{olt}) and (\ref{molt}) unchanged.
  In the special case that $K_{ij}$ is an everywhere nonzero multiple of the background metric
  $g_{ij}$, this new boundary condition is just a different way of writing the conformal
  boundary condition that we already studied in section \ref{works}, so in particular it is elliptic.   Lacking a better name, we will call what we get with eqn. (\ref{exc}) the
alternate boundary condition.  
  
As was remarked at the end of section (\ref{ellbc}),
  ellipticity is an ``open'' condition, preserved by any sufficiently small perturbation of a boundary condition.  Since the alternate boundary condition is elliptic
if  $K_{ij}$ is everywhere a nonzero multiple of $g_{ij}$, there must be an open set in the space of symmetric second rank tensors on $M$
 such that the alternate boundary condition is
  elliptic if $K_{ij}$ is everywhere in that open set.  
  
  To find this open set,  we can proceed almost as before.  We have to determine the large momentum behavior of a solution of $\D' h=0$ that looks like a plane
  wave along the boundary.  For this, we can take the usual flat model with  $X=\R^D_+$, $\partial X=\R^{D-1}$, and treat the tensor $K_{ij}$ that appears in the boundary condition (\ref{exc}) as a fixed 
  constant symmetric tensor.  Of course, in order for $K_{ij}$ really to be  the extrinsic curvature of the boundary, $X$ and its boundary cannot really be flat.  But their
  curvature does not affect the  high momentum behavior, which we can calculate using the flat model.
    
    Proceeding in this way,
  it is straightforward to compute that with the alternate boundary condition, eqns. (\ref{wolt}) and (\ref{noltt}) are replaced by
  \be\label{deffo}   \i \vec k\cdot \vec \alpha -\frac{1}{2}|\vec k|\beta +\frac{1}{2}|\vec k | \gamma \K=0\ee 
  \be\label{effo}-|\vec k|\alpha_i -\frac{\i}{2}k_i\beta +\i\gamma \left(k_j K_{ij}-\frac{1}{2}k_i\K\right) =0 .\ee
Likewise  eqn. (\ref{polot}) is replaced by
  \be\label{kiffo} 2\i \vec k\cdot \vec \alpha +|\vec k|\gamma \K=0. \ee
  Ellipticity is the statement that (at every point on $M=\partial X$) eqns. (\ref{deffo}), (\ref{effo}), and (\ref{kiffo}) have no common solutions with real nonzero $\vec k$.
  
  Comparing eqns. (\ref{deffo}) and (\ref{kiffo}), we see that a solution must have $\beta=0$, and once we know this, we can eliminate $\vec \alpha$ to find
  that a nonzero solution has  
  \be\label{num}\sum_{i,j}k_ik_j M_{ij}=0,\ee
   where $M_{ij}$ is the quadratic form
  \be\label{poff} M_{ij}=g_{ij} \K-K_{ij}. \ee
  The condition for eqn. (\ref{num}) to have no nonzero real solution is simply that the quadratic form $M_{ij}$ should be positive-definite or negative-definite.
  For this it is sufficient, though not necessary, that the extrinsic curvature $K_{ij}$ of the background metric should be positive- or negative-definite.
  
Now let us suppose that the quadratic form $M$ is positive- or negative-definite, so that the alternate boundary condition is elliptic.  What does this say about the linearized
Einstein equations with Dirichlet boundary conditions?

 In what follows, we will write $\D''$ for the gauge-fixed linearized Einstein operator with alternate boundary conditions and $\D'$ for the same operator with Dirichlet
 boundary conditions. 
If $\D''$ is elliptic, it has in particular a finite-dimensional cokernel.  This means
that, given a symmetric tensor $f$ on $X$, imposing finitely many linear constraints on $f$ suffices 
to ensure that the gauge-fixed equation
\be\label{noff}\D'' h = f \ee
has a solution, with $h$ obeying the alternate boundary conditions of eqns. (\ref{olt}), (\ref{molt}), and  (\ref{exc}).   The $h$ 
that satisfies these boundary conditions  does not in general satisfy Dirichlet boundary conditions, of course.   
According to eqn. (\ref{exc}), the Dirichlet boundary conditions are violated  because $h_{ij}|$, instead of vanishing, is instead
\be\label{proff} h_{ij}|= \gamma K_{ij}\ee
 for some function $\gamma$ on $M=\partial X$.   However, we can compensate for this by shifting
\be\label{roff} h_{\mu\nu}\to h_{\mu\nu}+D_\mu v_\nu +D_\nu v_\mu \ee
for a suitable vector field $v^\mu$.   We require first of all that 
\be\label{tfo} P v = 0, \ee
so that the shift (\ref{roff}) does not disturb the equation (\ref{noff}) or the gauge condition $T_\mu(h)=0$.   Second, we require that $v$ satisfies the boundary conditions
\be\label{lopp} v^i|=0,~~~ v^\perp|=\gamma , \ee analogously to what we did previously in eqn. (\ref{wyrd}).
Invertibility of $P$ on a manifold with boundary, as proved in Lemma 2.2 of \cite{Anderson}, means that such a $v$ exists.  (If this argument were not available,
we would say that ellipticity of $P$ implies that $v$ exists after possibly placing finitely many additional linear 
constraints on $\gamma$. This would be enough for what follows.)  Eqn. (\ref{lopp}), together with the fact that
$K_{ij}$ is the normal derivative to the metric of $M$, means that the shift (\ref{roff})  eliminates the right hand side of eqn. (\ref{proff}) and sets $h_{ij}|=0$.
Thus (as in Proposition 3.5 of \cite{Anderson2})
ellipticity of $\D''$ implies that after imposing finitely many constraints on $f$, the equation $\D' h=f$ can be satisfied with an $h$
that obeys Dirichlet boundary conditions; in other words, it implies that $\D'$ has  a finite-dimensional cokernel, as if it were elliptic.

Once we know that the cokernel of $\D'$ is finite-dimensional, it follows immediately from eqn. (\ref{welg}) that its kernel is
also finite-dimensional.  Suppose that $\D'\t h=0$ for some $\t h$ that satisfies Dirichlet boundary conditions.  
This implies that we cannot solve eqn. (\ref{noff}) unless $\la f,\t h\ra=0$, since
if we can solve eqn. (\ref{noff}), then
\be\label{hoft} \la f,\t h\ra =\la \D' h,\t h\ra =\la h, \D' \t h\ra = 0. \ee
Thus every element of the kernel of $\D'$ gives a constraint on $f$, so the dimension of the kernel  of $\D'$ can be no greater than the dimension of the cokernel,
and in particular  the kernel has finite dimension if the cokernel does.  

\def\H{{\mathcal H}}
Actually, we can be more precise here.  Let us think of the space of all metric perturbations as a Hilbert space $\H$ with inner product $\la~,~\ra$.
Once we know that the cokernel of $\D'$  is finite-dimensional, it follows\footnote{By constrast, if
 $V$ is a linear subspace of $\H$ that is not of finite codimension, then in general
$V$ is not a Hilbert space and $\H/V$ cannot be naturally identified with $V^\perp$. Rather, $V^\perp$ can be identified with $\H/\overline V$, where
$\overline V$ is the Hilbert space closure of $V$.}   that the image
of $\D'$ is a Hilbert subspace $\H'\subset \H$  and that the cokernel of $\D'$, which is $\H/\H'$,
 can be identified with the orthocomplement of $\H'$.    Thus, we can identify the cokernel of
$\D'$ with the
space of all metric perturbations $\t h$ obeying Dirichlet boundary conditions that are orthogonal to $\D' h$ for any $h$ that obeys Dirichlet boundary conditions.
But this orthogonality together with eqn. (\ref{welg}) gives
\be\label{noft} 0 =\la \D' h, \t h \ra = \la h, \D' \t h\ra. \ee
Since this is supposed to be true for all $h$ that obey Dirichlet boundary conditions, it follows that $\D' \t h=0$.  We can also read this backwards to show
that if $\D' \t h=0$, then $\t h$ is orthogonal to the image of $\D'$. Thus the kernel of $\D'$ is the orthocomplement of $\H'$ and so is isomorphic to the cokernel of $L'$.

All this is as if $\D'$ were elliptic when $\D''$ is elliptic.   That is certainly not true, since the failure of ellipticity of $\D'$ is universal. But it seems plausible (though apparently not known) 
that when $\D''$ is elliptic, $\D'$ has the necessary properties for perturbation theory
-- notably the existence of a suitable parametrix or propagator.
It seems doubtful that $\D'$  has reasonable properties in general, without ellipticity of the alternate boundary condition.   But little seems to be known
about this.

\vskip1cm
Research supported in part by NSF Grant PHY-1606531.    I thank R. Mazzeo and especially M. T. Anderson
for  advice about boundary conditions in gravity and helpful suggestions,  
and   M. J. Duff,  G. Horowitz, J. M. Maldacena, and D. Stanford for a variety of discussions.

\bibliographystyle{unsrt}

\begin{thebibliography}{99}
\bibitem{Anderson}
M. T. Anderson, ``Boundary Value Problems For Einstein Metrics,''  Geom. and Top. {\bf 12} (2008) 2009-45, arXiv:math/0612647.

\bibitem{Anderson2}
M. T. Anderson, ``Extension of Symmetries on Einstein Manifolds With Boundary,'' Selecta Math. {\bf 16} (2010) 343-375, arXiv:0704.3373.

\bibitem{AE}
I. G. Avramiki and G. Esposito, ``Lack Of Strong Ellipticity In Euclidean Quantum Gravity,'' 
Class. Quant. Grav. {\bf 15} (1998) 1141-1152,
arXiv:hep-th/9708163.

\bibitem{Hamilton}
R. S. Hamilton, ``The Inverse Function Theorem of Nash and Moser,'' Bull. Am. Math. Soc. {\bf 7} (1982) 65-222.

\bibitem{GH}
G. Gibbons and S. W. Hawking, ``Action Integrals And Partition Functions In Quantum Gravity,'' 
Phys. Rev. {\bf D15} (1977) 2752-6.

\bibitem{GP}
G. Gibbons and M. Perry, ``Quantizing Gravitational Instantons,'' Nucl. Phys. {\bf B146} (1978) 90.



\bibitem{GHP}
G. W. Gibbons, S. W. Hawking, and M. Perry, ``Path Integrals and the Indefiniteness of the Gravitational Action,'' Nucl. Phys. {\bf B138} 
(1978) 141-50.

\bibitem{CD}
S. M. Christensen and M. J. Duff, ``Quantizing Gravity With A Cosmological Constant,'' Nucl. Phys. {\bf B170[FS1]} (1980) 480-506.



\bibitem{HH}
J. Hartle and S. W. Hawking, ``The Wavefunction Of The Universe,'' Phys. Rev. {\bf D28} (1983) 2960-75.

\bibitem{Polch} J. Polchinski, ``The Phase Of The Sum Over Spheres,'' Phys. Lett. {\bf B219} (1989) 251-7.

\bibitem{AEK}
I. G. Avramidi, G. Esposito, A. Yu. Kamenshchik, ``Boundary Operators In Euclidean Quantum Gravity,'' 
Class. Quant. Grav. {\bf 13} (1996) 2361-2374,
arXiv:hep-th/9603021.

\bibitem{BB}
F. Bastianelli and R. Bonezzi, ``One-Loop Quantum Gravity From A Worldline Viewpoint,'' arXiv:1304.7135.


\bibitem{York}
J. W. York, ``Role Of Conformal Three-Geometry In The Dynamics Of Gravitation,'' Phys. Rev. Lett. {\bf 28} (1972) 1082.

\bibitem{Talk}
E. Witten, ``Canonical Quantization In Anti de Sitter Space,'' lecture available at \url{http://pcts.princeton.edu/pcts/20YearsAdSCFT/slides+videos.html}.

\bibitem{CCJ}
S. Coleman, C. G. Callan, Jr., and R. Jackiw, ``A New Improved Energy-Momentum Tensor,'' Ann. Phys. {\bf 59} (1970) 42-73.

\bibitem{Anderson3}
M. T. Anderson, ``On Quasi-Local Hamiltonians in General Relativity,'' Phys. Rev. {\bf D82} (2010) 084044,
arXiv:1008.4309.

\end{thebibliography}

\end{document}